\documentclass{ws-mpla}

\usepackage[final]{showkeys}

\begin{document}

\markboth{Peris, Boito, Golterman and Maltman}
{The case for Duality Violations in the analysis of hadronic $\tau$ decays}

\catchline{}{}{}{}{}

\title{THE CASE FOR DUALITY VIOLATIONS \\
IN THE ANALYSIS OF HADRONIC $\tau$ DECAYS
}

\author{SANTIAGO PERIS\footnote{Speaker.}}

\address{Department of Physics and IFAE-BIST, Univ. Autonoma de Barcelona \\
E-08193 Bellaterra, Barcelona, Spain\\
peris@ifae.es
}

\author{DIOGO BOITO}
\address{S\~ao Carlos Institute of Physics, University of S\~ao Paulo\\
PO Box 369, 13570-970, S\~ao Carlos, SP, Brazil}

\author{MAARTEN GOLTERMAN}

\address{Department of Physics and Astronomy, San Francisco State Univ.,\\
San Francisco, CA 94132, USA}

\author{KIM MALTMAN}

\address{Department of Mathematics and Statistics, York Univ. \\
Toronto, Ontario M3J 1P3, Canada\\
CSSM, Univ. of Adelaide, Adelaide, South Australia 5005, Australia}

\maketitle

\pub{Received (Day Month Year)}{Revised (Day Month Year)}

\begin{abstract}

We discuss why, in the determination of $\alpha_s(m_{\tau}^2)$ from hadronic $\tau$ decays, two important assumptions made in most of previous analyses, namely the neglect of higher-dimension condensates and of Duality Violations (DVs), have introduced uncontrolled systematic errors into this determination. Although the use of pinched weights is usually offered as a justification of these assumptions, we explain why it is not possible to simultaneously suppress  these two contributions; particularly since the Operator Product Expansion is expected to be an asymptotic, rather than a convergent expansion. There is not only experimental and theoretical evidence for DVs, but they also affect the extraction of $\alpha_s(m_{\tau}^2)$.

\keywords{QCD; $\tau$ decays; $\alpha_s$.}
\end{abstract}

\ccode{PACS Nos.: include PACS Nos.}

\section{Introduction}	

In 1992, Braaten, Narison and Pich,\cite{BNP} and later on Le Diberder and Pich,\cite{oldstrategy} culminated a series of previous investigations\cite{Shankar}$^{-}$\cite{Braaten88} on the hadronic decay of the $\tau$ lepton with an analysis which revealed that this decay is a very interesting one for studying QCD. This led to, in particular, a very competitive  determination of $\alpha_s$; a determination which has survived until now. However, today, 24 years later, the increase in precision of the experimental data\cite{Davier} and the improved theoretical knowledge\cite{Shifman0} has called for a thorough reinvestigation of this type of analysis, which will be discussed in more detail in section 5, and to which we will refer as ``the old-strategy analysis."  This reinvestigation of the old-strategy analysis has exposed as two of the main sources of systematic errors the neglect of the contribution from higher-order condensates and the neglect of the contributions from the so-called quark-hadron Duality Violations(DVs).\cite{Shifman0} As we will discuss below, these two contributions turn out to be related.

 DVs refer to the failure of the Operator Product Expansion (OPE) to properly describe the physics on the Minkowski axis and, in particular, to reproduce the spectral functions which are the primary experimental data. Our modification of the old strategy, consisting of a series of papers dealing with several aspects of the analysis,\cite{Cata0}$^{-}$\cite{Boito3} has culminated in the work of Ref.~\refcite{us} which has produced, among other things, a new value for $\alpha_s(m_{\tau}^2)$. This is where we currently stand. A detailed account of this new analysis will be presented in these proceedings by D. Boito.\cite{Boito0}  In this article, on the other hand, we will focus on the connection between the contribution from higher-dimension condensates and DVs, and their subsequent impact on the determination of $\alpha_s(m_{\tau}^2)$.

The $V_{\mu}=\overline{u}\gamma_{\mu} d$ and  $A_{\mu}=\overline{u}\gamma_\mu\gamma_5 d$ spectral functions with $u,d$ quark content, $ \rho_{V/A;ud}(s)$,  can be
experimentally determined from the ratio
\begin{equation}
\label{R}
R_{V/A;ud}=
{\frac{\Gamma [\tau\rightarrow ({\rm hadrons})_{V/A;ud}\nu_\tau (\gamma ) ]}
{\Gamma [\tau\rightarrow e\bar{\nu}_e \nu_\tau (\gamma ) ]}}\ ,
\end{equation}
as,\cite{Tsai}
\begin{equation}
\label{taukinspectral}
{\frac{dR_{V/A;ud}(s)}{ds}}= \frac{12\pi^2\vert V_{ud}\vert^2 S_{EW}}{m_\tau^2}\,
 \left[ w_T(s/m_\tau^2) \rho_{V/A;ud}^{(1+0)}(s)
- w_L(s/m_\tau^2) \rho_{V/A;ud}^{(0)}(s) \right]\ ,
\end{equation}
where $S_{EW}$ is a short-distance electroweak correction and
$w_T(x)=(1-x)^2(1+2 x)$, $w_L(x)=2 x (1-x)^2$ are polynomials with a double zero at $x=1$. This is referred to in the usual jargon as ``doubly pinched." As we will see, the use of pinching will be important in the discussion that follows. The spectral functions  $\rho^{(1,0)}(s)$ are related to the corresponding two-point current correlators $\Pi^{(1,0)}_{\mu\nu}(q)$ as $\rho^{(1,0)}(s)=(1/\pi) \mathrm{Im}\Pi^{(1,0)}(s)$, where the scalar functions $\Pi^{(1,0)}(q^2)$ are defined as
\begin{eqnarray}
\label{correl}
\Pi_{\mu\nu}(q)&=&i\int d^4x\,e^{iqx}\langle 0|T\left\{J_\mu(x)J^\dagger_\nu(0)\right\}|0\rangle\\
&=&\left(q_\mu q_\nu-q^2 g_{\mu\nu}\right)\Pi^{(1)}(q^2)+q_\mu q_\nu\Pi^{(0)}(q^2)\nonumber\\
&=&\left(q_\mu q_\nu-q^2 g_{\mu\nu}\right)\Pi^{(1+0)}(q^2)+q^2 g_{\mu\nu}\Pi^{(0)}(q^2)\ .\nonumber
\end{eqnarray}
$J_\mu$ stands for the non-strange $V$ or $A$ current,
$\overline{u}\gamma_\mu d$ or $\overline{u}\gamma_\mu\gamma_5 d$, while
the superscripts $(0)$ and $(1)$ label spin.  In what follows we will always take the $0+1$ spin combination because it is free from kinematical singularities.
\begin{figure}[t]
\centerline{\psfig{file=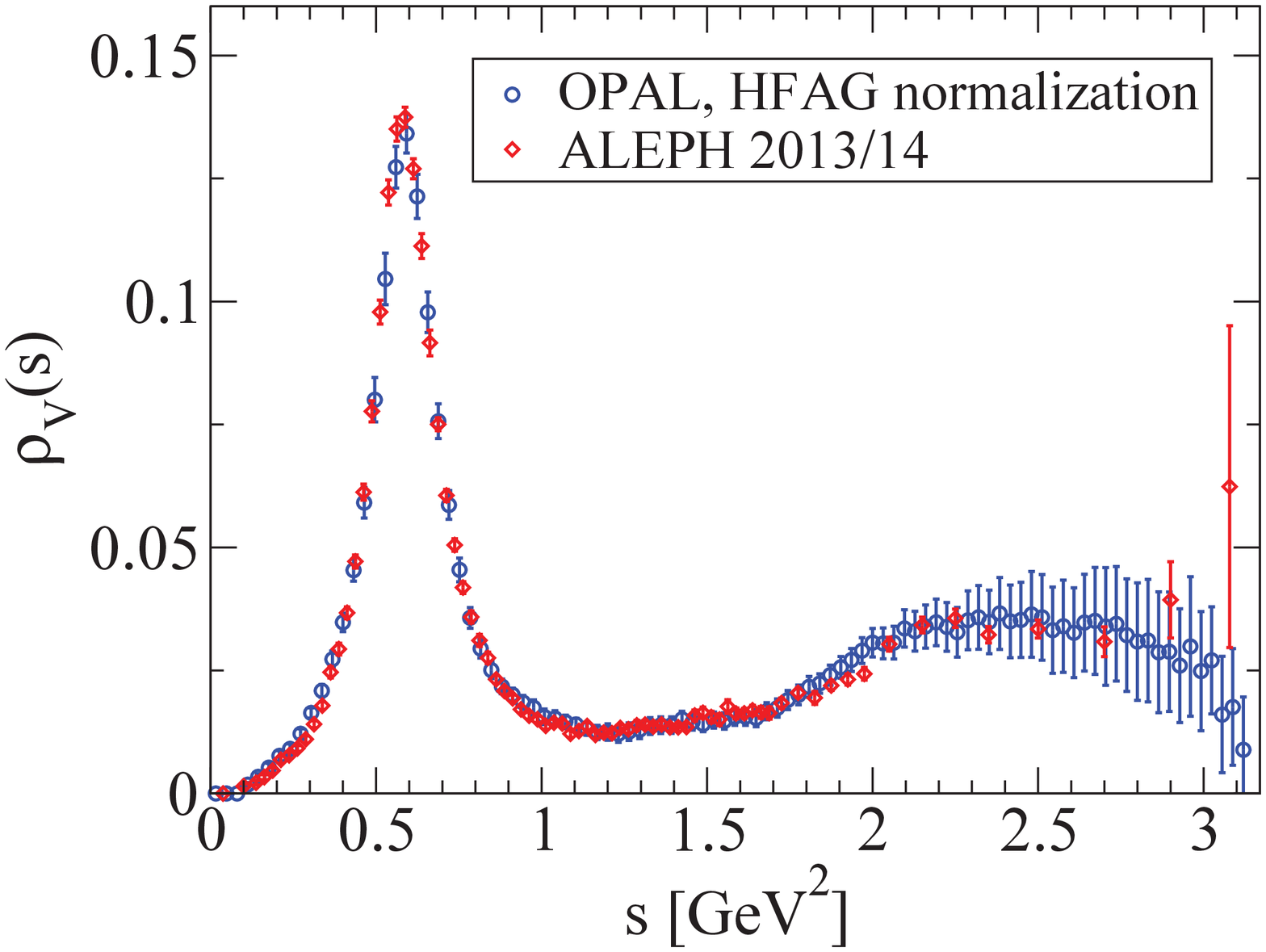,width=2.2in}\psfig{file=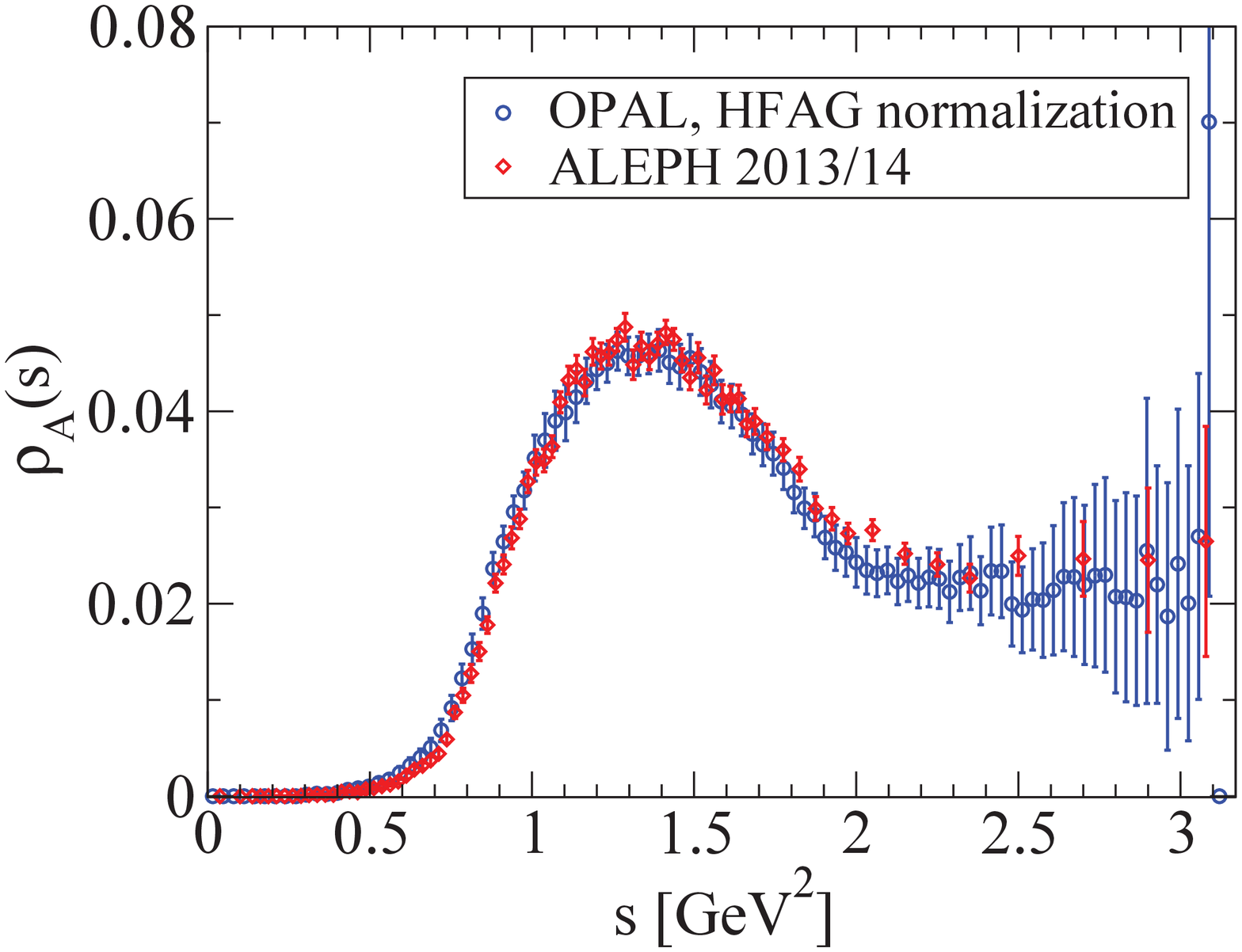,width=2.2in}}
\vspace*{8pt}
\caption{Opal and Aleph spectral functions $\rho^{(1+0)}_{V/A}(s)$.\protect\label{fig1}}
\end{figure}
We are using the notation $s=q^2$, so that $q^2>0$ in the Minkowski.

The spectral function $\rho^{(0)}_{V}(s)$ is negligible and  $\rho^{(0)}_{A}(s)$ is dominated by the pion pole, which is known. The combination $\rho^{(1+0)}_{V/A}(s)$ may, therefore, be determined directly
from $dR_{V/A;ud}(s)/ds$. This provides the primary experimental data the following analysis will be based on, and it is shown in Fig.~\ref{fig1}.\cite{Davier,Opal} As one can see, even above $s\simeq 1.5\, \mathrm{GeV}^2$, the data shows clear oscillations not present in the OPE prediction. The OPE prediction, i.e., perturbation theory plus condensates, in fact, corresponds to a nearly horizontal line with a value of $\sim 0.025$, in clear disagreement with the experimental data.

\section{Theoretical Foundations}	

\begin{figure}[t]
\centerline{\psfig{file=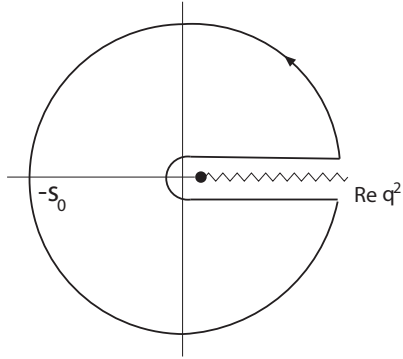,width=2.2in}}
\vspace*{8pt}
\caption{Contour used in Eq.~\ref{cauchy}.\protect\label{fig2}}
\end{figure}

 Even though the OPE locally disagrees with the spectral data, one expects that at least globally there might be a better agreement. Therefore, one considers Finite Energy Sum rules (FESRs) i.e.,  integrals of the form
 \begin{eqnarray}
\label{cauchy}
I^{(w)}_{V/A}(s_0)\equiv\frac{1}{s_0}\int_0^{s_0}ds\,w(s)\,\rho_{V/A}(s)
&=&-\frac{1}{2\pi i\, s_0}\oint_{|s|=s_0}
ds\,w(s)\,\Pi_{V/A}(s)\ ,
\end{eqnarray}
valid for any $s_0>0$ and any weight $w(s)$ analytic inside and on the
contour represented in Fig.~\ref{fig2}.\cite{BNP,Shankar}$^{-}$\cite{Braaten88} In the second equality, Cauchy's theorem has been used on account of the analytic structure of the correlators $\Pi(q^2)$, which have a cut on the positive real axis,\footnote{The axial channel also has the pion pole.} precisely where the spectral functions are measured. As it stands, Eq.~(\ref{cauchy}) is an exact mathematical statement as long as the exact correlator $\Pi(q^2)$ is used. However, this correlator is not known exactly from QCD. If the scale $s_0$ is deemed ``large enough" (more about the perils of this assumption in the following), it begins to make sense to use of the OPE on the contour and write the right-hand side of Eq.~(\ref{cauchy}) as
\begin{equation}
\label{cauchyDV}
=-\frac{1}{2\pi i\, s_0}\oint_{|s|=s_0}
ds\,w(s)\,\left[\Pi^{OPE}_{V/A}(s)+  \Pi^{DV}_{V/A}(s)\right]\ ,
\end{equation}
where, by definition, $\Pi^{DV}_{V/A}(s)=\Pi_{V/A}(s)-\Pi^{OPE}_{V/A}(s)$. It is precisely the term $\Pi^{DV}_{V/A}(s)$ which will be the subject of the following discussion. The superscript ``DV" stands for \emph{Duality Violations} (DVs) and refers to the fact that the correlator computed with the OPE in terms of quark and gluon fields does not equal the hadronic counterpart. Apart from the term encoding DVs, Eqs.~(\ref{cauchy}) and (\ref{cauchyDV}) yield the necessary relation to determine QCD parameters (such as $\alpha_s$, contained in $\Pi^{OPE}_{V/A}$), from the experimental data (contained in $\rho_{V/A}$).

If there were a limit in which $\Pi^{OPE}_{V/A}(s) \rightarrow \Pi_{V/A}(s)$ as, e.g., in a convergent expansion, then one would of course have $\Pi^{DV}_{V/A}(s) \rightarrow 0$. However, there is rather compelling evidence that the OPE is \emph{not} a convergent expansion.\footnote{Although, to the best of our knowledge, no formal proof is available.} Perhaps the simplest way to see this is by considering the OPE as an expansion in $1/q^2$, i.e. neglecting all the logarithmic corrections. In this case the OPE becomes a power expansion and, as such, its assumed convergence should happen in a full disc in the complex plane around $1/q^2=0$ . Clearly, the existence of the cut shown in Fig.~\ref{fig2}, running all the way  to infinity, contradicts this. The large-$N_c$ limit also suggests the non-convergence of the OPE. In the large-$N_c$ limit the spectral function has poles at arbitrarily high positive $q^2$ (i.e., the meson masses). Obviously there is no sense in which $q^2$ is ``larger than any meson mass" to make a large-$q^2$ expansion a convergent  one. A somewhat more elaborate argument can also be found in Ref.~\refcite{Shifman1}.

 If the OPE is not convergent, there is no sense in which $s_0$ may be ``large enough" to guarantee that $\Pi^{DV}\sim 0$ in Eq.~(\ref{cauchyDV}). In the case of the physical decay of the $\tau$ lepton, the scale $s_0= m_{\tau}^2$. So, the argument that $m_{\tau}$ is a large scale suppressing the contribution of high-dimension condensates in the OPE, as has been invoked in the literature in the past, is unreliable.

But not everything is lost. If the OPE is not convergent, the next logical possibility is that it is at least asymptotic. Indeed,  asymptotic expansions may have the property of approaching the true function only in an angular sector of the complex plane. The fact that the OPE does a good job describing the correlator in the Euclidean ($q^2<0$) while failing in the Minkowski where the cut is located ($q^2>0$) would suggest this is the case. This would come at a price, however. Unlike convergent expansions, an asymptotic expansion misses a term relative to the true function.  This missing term  is of order $\mathrm{e}^{-\gamma/\alpha}$ for the case of an expansion parameter $\alpha>0$. The constant $\gamma$ depends on the case considered. In the case of renormalons, for instance, one has $\alpha=\alpha_s$ and $\gamma$ is related to (the inverse of) the beta function. Therefore, in the case of the OPE, since the expansion parameter is $1/q^2$, one should expect a correction on the Minkowski axis like $\Pi^{DV}\sim \mathrm{e}^{-\gamma q^2}$, for $q^2>0$ and some unknown parameter $\gamma$.

This is when ``pinching" enters our discussion. If the weight function $w(s)$ in Eq.~(\ref{cauchyDV}) is chosen with a high-order zero at $s=s_0$, the contribution from the contour in the region around the Minkowski axis, where $\Pi^{DV}$ has its support, will be suppressed. Pinching, therefore, comes as a potentially useful trick to suppress DVs. Although this may be true qualitatively, at a more quantitative level it is far from clear how much of a suppression one really has for a given order in the pinched polynomial. This is of course crucial when good control over the systematics is needed in a reliable determination of an important parameter of QCD such as~$\alpha_s$.

The above discussion leads to the main theoretical message of this article:
\vspace{.5cm}

\centerline{``\emph{It is not possible to simultaneously suppress DVs and higher-dimension condensates.}"}

\vspace{.5cm}

In other words: pinching comes at a price. Indeed, a high-order zero at $s=s_0$ in the polynomial $w(s)$ is only achieved when $w(s)$ is a high-order polynomial. However, since a term of order $s^n$ selects a condensate of dimension $2(n+1)$ in Eq.~(\ref{cauchyDV}), a high-order polynomial also selects high-order condensates, which are typically either poorly known or unknown.

If the OPE were a convergent expansion, this would be a minor problem as higher-dimension condensates would contribute less and less. This, in fact, has been the main (explicit or tacit) assumption in most of the literature so far. As the OPE is expected to be an asymptotic expansion, neglecting higher-dimension condensates becomes a very crude approximation and, at the very least,   potentially introduces uncontrolled systematics one should avoid in any case.

\section{Duality Violations: Generalities}

Up to now, there is no theory of DVs in QCD from first principles. So, a quantitative analysis of DVs faces serious theoretical challenges and it is no surprise that DVs have been neglected in the old-strategy analysis of hadronic $\tau$ decays. Nevertheless, some progress can be made.

\begin{figure}[t]
\centerline{\psfig{file=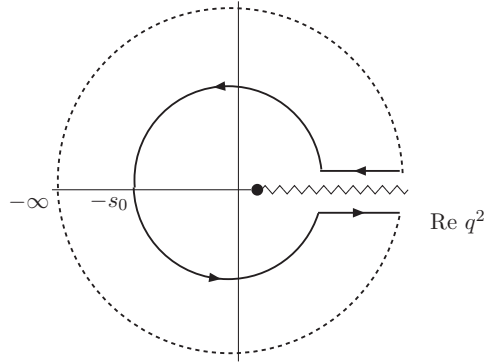,width=2.5in}}
\vspace*{8pt}
\caption{Contour used to obtain Eq.~(\ref{cauchyreverse}).\protect\label{fig3}}
\end{figure}

First, a generic property of an asymptotic expansion is that it exactly agrees with the true function only for a vanishing value of the  expansion parameter. This assures that $\Pi^{DV}(s)\rightarrow 0$ as $s\rightarrow \infty$ (in any direction in the complex plane). Then, using Cauchy's theorem ``in reverse" by closing the contour at infinity (which gives a vanishing contribution) as depicted in Fig.~\ref{fig3}, one obtains

\begin{equation}
\label{cauchyreverse}
-\frac{1}{2\pi i}\oint_{|z|=s_0} dz\  w(z)\  \Pi^{V/A}_{DV}(z)= \  - \int_{s_0}^{\infty}\!\!\!ds\  w(s)\ \   \frac{1}{\pi}\mathrm{Im}\Pi^{V/A}_{DV}(s)\ .
\end{equation}

The right-hand side of Eq.~\ref{cauchyreverse} makes explicit two important properties of DVs. First, the contribution from DV to FESRs involves an extrapolation from $s_0$ to infinity, a region which is not covered by experimental $\tau$ data. Second, recalling that  DVs are exponentially damped, one clearly sees why a zero of the weight $w(s)$ at $s=s_0$, i.e. the lower end of the integration interval,  tends to suppress the contribution of the whole integral.\footnote{For a more detailed discussion of this suppression, see section 7.}

\section{Duality Violations: An educated guess}

\begin{figure}[t]
\centerline{\psfig{file=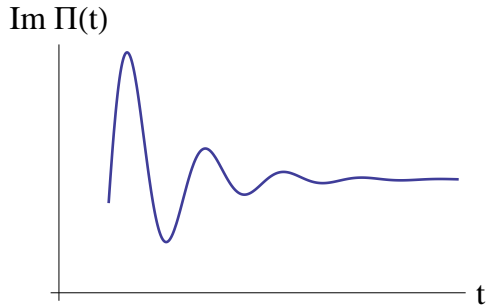,width=2.5in}}
\vspace*{8pt}
\caption{Schematic representation of the spectrum in Fig.~\ref{fig1}, showing DVs.\protect\label{fig4}}
\end{figure}

Without an explicit representation of DVs it is not possible to go further. We need a concrete parametrization. Keeping in mind the expected qualitative behavior shown in Fig.~\ref{fig4} and the previous discussion, we find it natural to assume\cite{Cata0}
\begin{equation}
\label{DV}
\frac{1}{\pi}\mathrm{Im}\Pi^{V/A}_{DV}(s) \simeq{ \mathrm{e}^{-\delta_{V/A}}\ \mathrm{e}^{-\gamma_{V/A}\,s}\  \sin(\alpha_ {V/A} +\beta_{V/A}  s)}
\end{equation}
as the ansatz for the contribution of DVs in Eq.~(\ref{cauchyreverse}), for large eonugh $s$. The exponential damping in $s$ stems from the asymptotic property of the OPE, as discussed above. The oscillation is assumed periodic with frequency $\beta_{V/A}$ as inspired by Regge theory. It is clear that the crests seen in Fig.~\ref{fig4} are due to the different resonances,  which would become Dirac deltas in the large-$N_c$ limit. The phenomenological success of Regge theory predicting daughter trajectories i.e., an equal spacing rule for the squares of the resonance masses, suggests a periodic function with a single frequency as a reasonable first approximation. The amplitude of the oscillation is parametrized as $\mathrm{e}^{-\delta_{V/A}}$ just for numerical convenience. We emphasize that, although the functional form is the same for $V$ and $A$, the values of the parameters may differ since  the two spectra approach the OPE result differently. This adds 4+4=8 parameters in total.

Let us conclude this section by noting  that a parametrization of the form of Eq.~(\ref{DV}) can be explicitly shown to be true in a specific model constructed to study DVs in Ref.~\refcite{Blok}, and has been applied to determine low-energy constants and condensates in QCD from the $V-A$ correlator by different groups in Refs.~\refcite{Gonzalez}-\refcite{Rodriguez}.

\section{Critical Review of the ``Old-Strategy" Method}

The old-strategy method\cite{oldstrategy} consists in using five  pinched weights $w_{kl}(y)=(1-y)^2(1+2y)(1-y)^k y^l$, for $(k,l)=\{(0,0), (1,0), (1,1),(1,2), (1,3)\}$, with $y=s/s_0$, in the FESRs (\ref{cauchy}-\ref{cauchyDV}) for a  \emph{single} value of $s_0=m_{\tau}^2$. We emphasize that these polynomials go up to $s^7$ which, through the FESR (\ref{cauchyDV}), requires knowledge of the condensates up to dimension 16. However, only condensates of dimension $4,6$ and $8$ are kept. Condensates of higher dimension, $10,12,14$ and $16$, are  set to zero by fiat. The argument behind this assumption is that the relevant scale $s_0=m_{\tau}^2$ is large enough, and the OPE is convergent enough, to make the contribution of these higher order condensates numerically negligible. However, as we have seen, since the OPE is asymptotic this assumption is dangerous. Furthermore, DVs are also set to zero on the basis of the above weights being doubly and triply pinched. The danger of employing a high degree of pinching (and hence a high degree polynomial) while at the same time neglecting the higher dimension OPE contributions introduced by the use of the higher degree polynomial, is precisely the point of our main theoretical message expressed in Section 2. As we will see, although DVs are indeed suppressed for these weights, they are not totally negligible. In fact, although pinching suppresses DVs, the \emph{amount} of this suppression is not necessarily correlated with the amount of pinching \emph{when a single $s_0$ is employed}, as in the old-strategy analysis. That is, e.g., a triply pinched weight may have a larger DV contribution than a doubly pinched weight when only a single $s_0$ value such as $s_0=m_{\tau}^2$ is considered (see section 7).

Given the five moments constructed with the five pinched weights $w_{kl}$ defined above, a fit is performed to extract 4 parameters: the condensates of dimension $4,6$ and $8$, and $\alpha_s(m_{\tau}^2)$. This is a fit with only one degree of freedom. The results of this fit\cite{Davier} reveal an inconsistency in the extracted value of the gluon condensate, where in the $V$ channel one has $\langle\frac{\alpha_s}{\pi}GG\rangle=(-0.5\pm 0.3)\times
10^{-2}~\mbox{GeV}^4,  (\chi^2=0.43, \ p=51\%)$, while in the $V+A$ channel one has $(-2.0\pm 0.3)\times 10^{-2}~\mbox{GeV}^4\ , (\chi^2=1.1, \ p=29\%)$, which are incompatible. Clearly the gluon condensate should have come out to be the same. The $A$ channel produces a third incompatible number. Although the $V$ channel result is the one statistically preferred (i.e., a higher p value), it is the $V+A$ channel's result which is chosen for the extraction of QCD parameters and, in particular, of $\alpha_s$.\cite{Davier}

\begin{figure}[t]
\centerline{\psfig{file=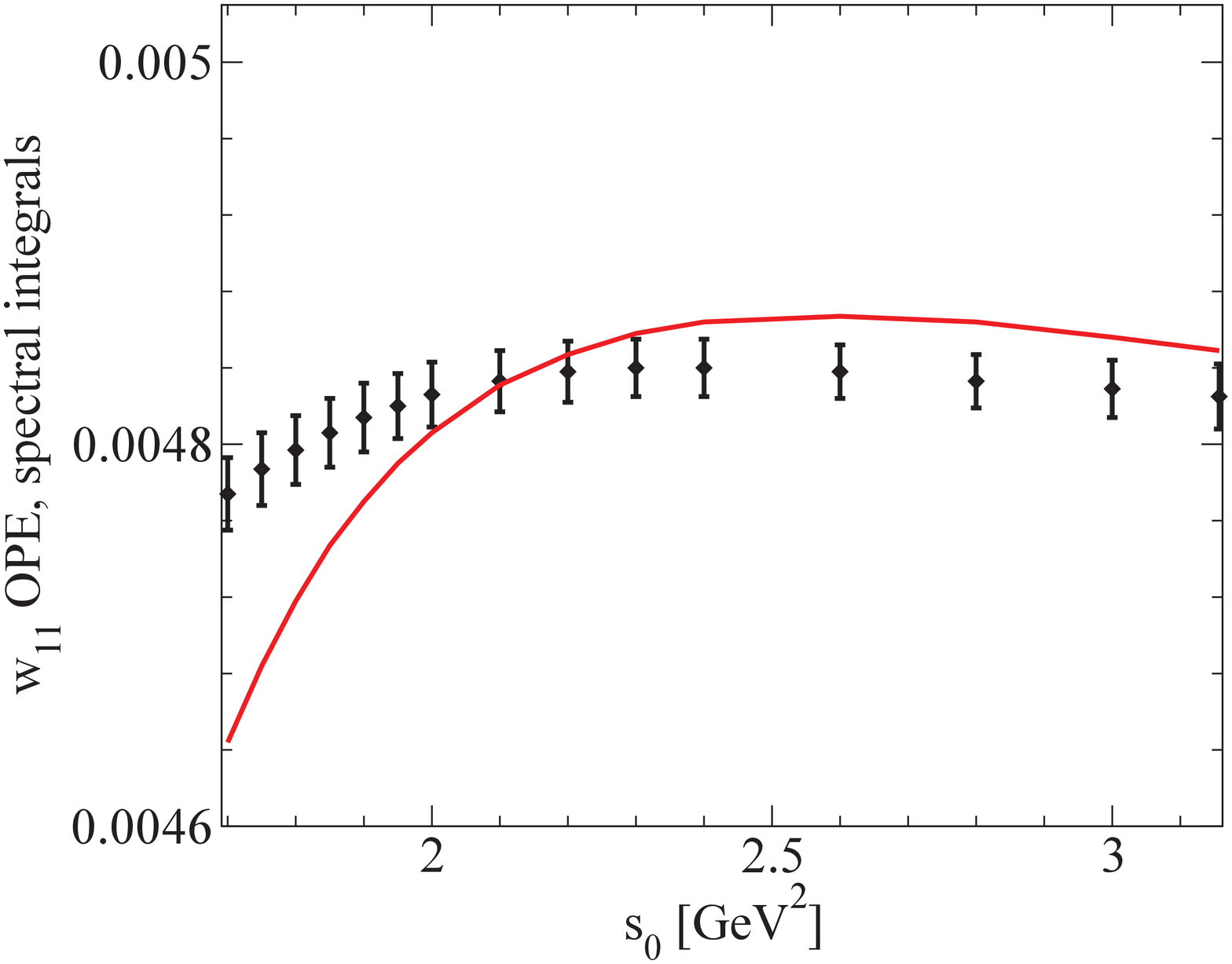,width=2.5in}
\psfig{file=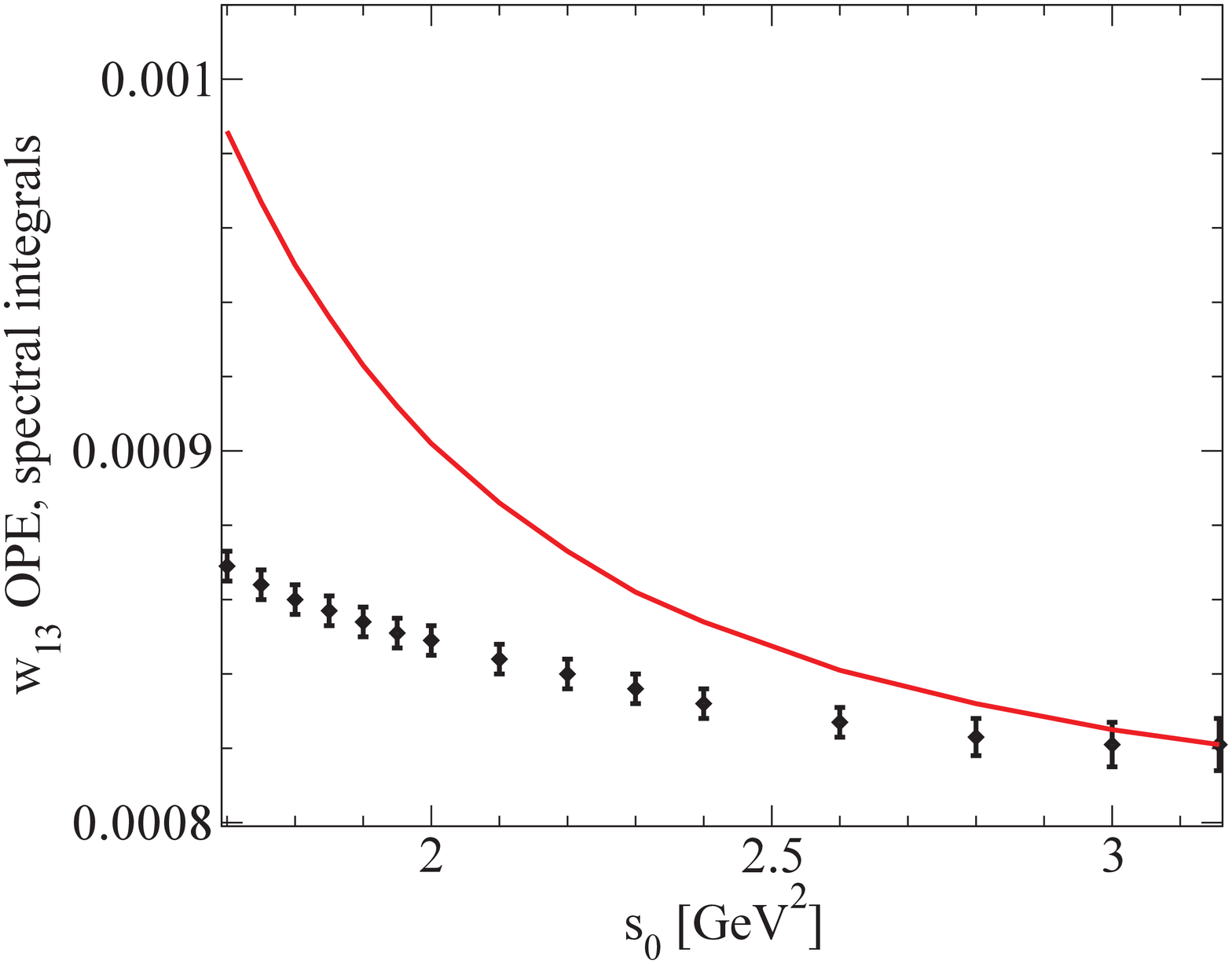,width=2.5in}}
\vspace*{8pt}
\caption{$s_0$ dependence of the $V+A$  $w_{11}$ and $w_{13}$ FESR  for CIPT from Ref.~\protect\refcite{Davier}.\label{fig5}}
\end{figure}

\section{Comparing tests}

\begin{figure}[t]
\centerline{\psfig{file=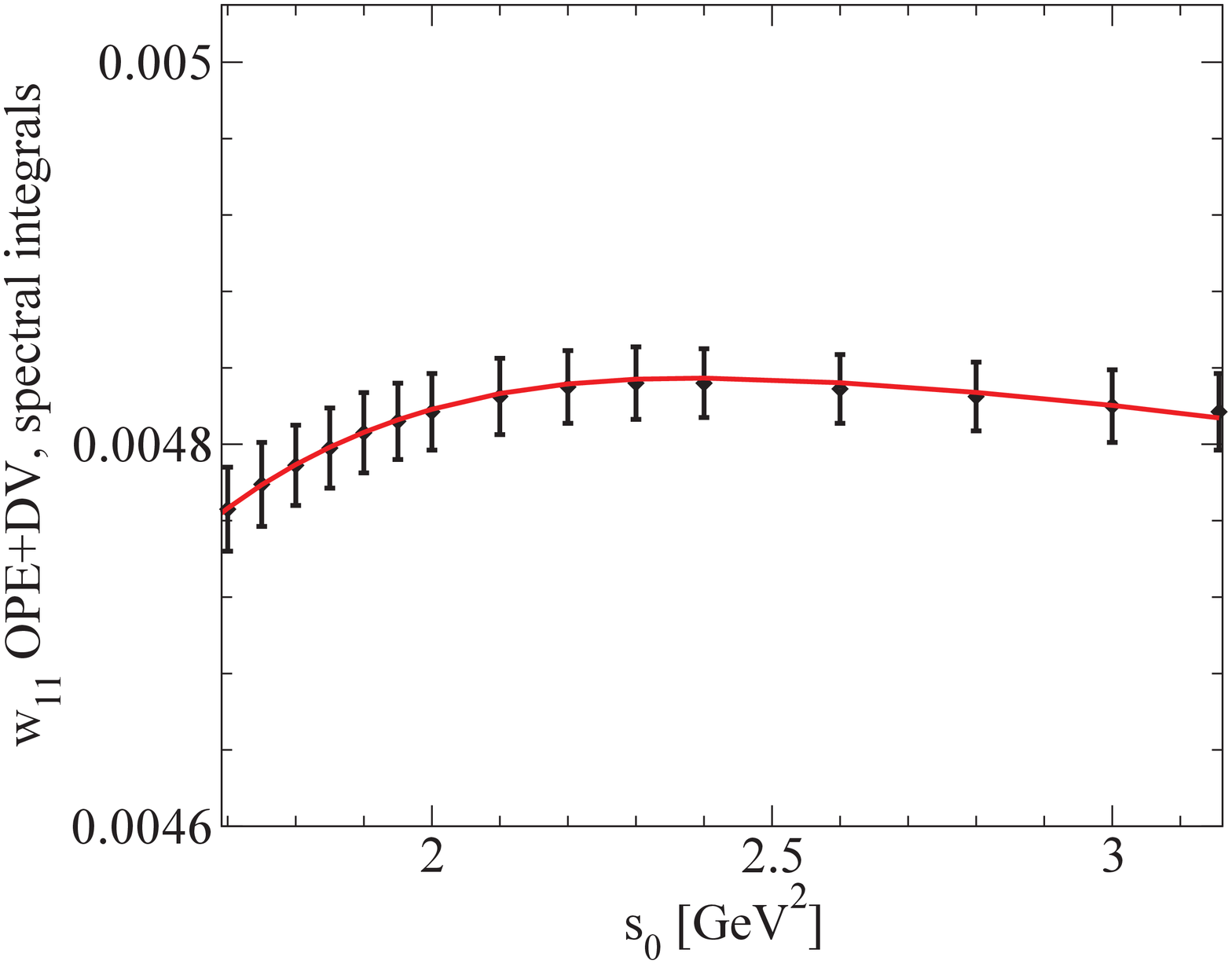,width=2.5in}
\psfig{file=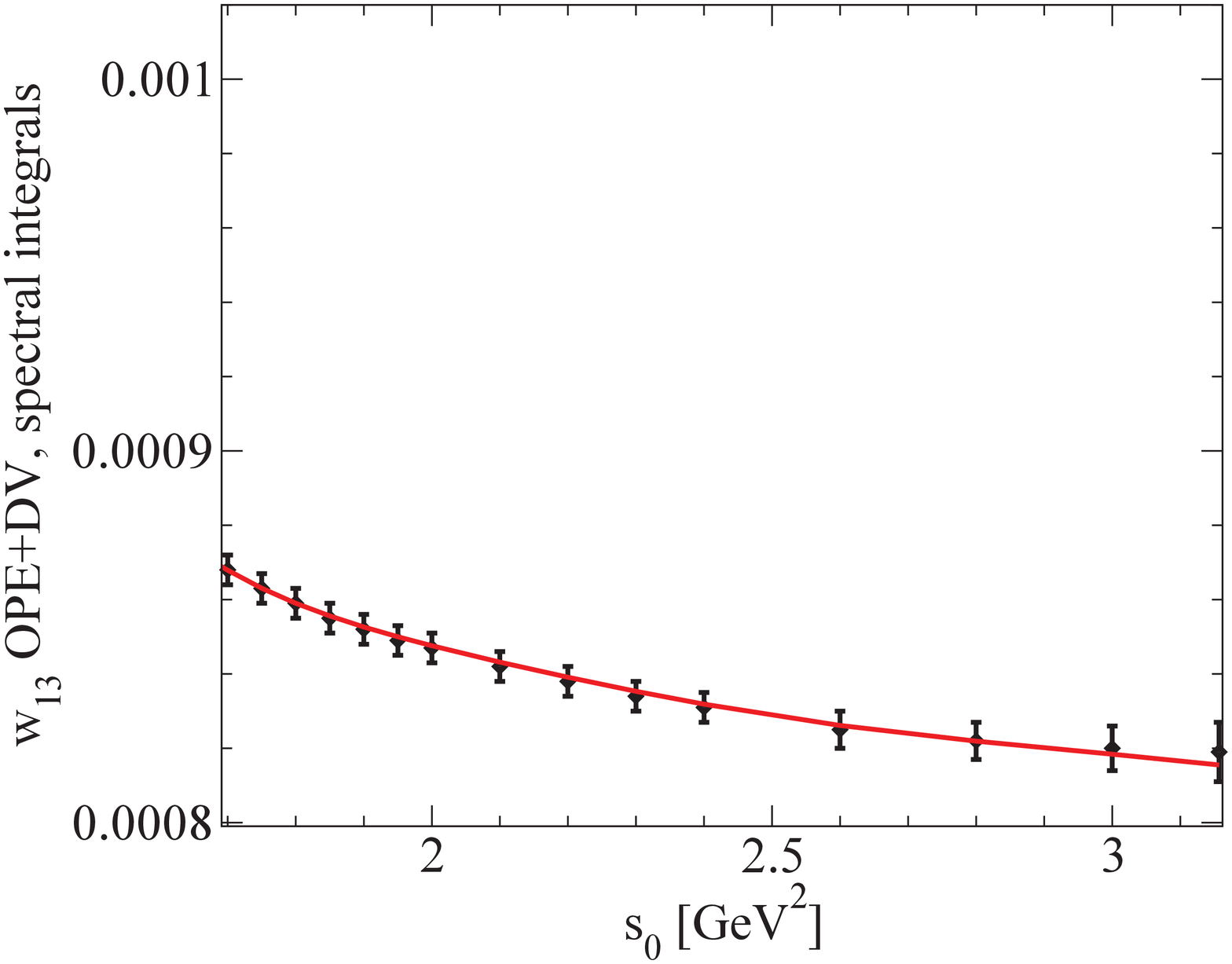,width=2.5in}}
\vspace*{8pt}
\caption{$s_0$ dependence of the $V+A$  $w_{11}$ and $w_{13}$ FESR  for CIPT from Ref.~\protect\refcite{us}.\label{fig6}}
\end{figure}

Restricting the analysis to just $s_0=m_{\tau}^2$ is dangerous.\cite{maltmanyavin} With a fixed value of $s_0$ there is no way to determine whether neglected, but in fact non-negligible, higher-dimension condensates contaminate the results. The parameters will always do ``the best they can" to fit the data at this $s_0$. Because the contribution from different condensates scales differently with $s_0$, a nontrivial check is then whether these parameters can also describe the data when $s_0$ is lowered by some amount. If the result of the fit deviates from the data as soon as $s_0$ is lowered, this is a clear evidence of contamination by unaccounted for effects.

Fig.~\ref{fig5} shows the comparison of the result of the fit in Ref.~\refcite{Davier} and the data for two of the weight functions used in their fit: $w_{11}$ and $w_{13}$. As one can see, the fit representations deviate from the data as soon as $s_0$ is lowered.  For comparison, Fig.~\ref{fig6} shows the same plots but using the results of the fits obtained in Ref.~\refcite{us}. The difference is striking.


\section{Demystifying pinched weights}

\begin{figure}[t]
\centerline{\psfig{file=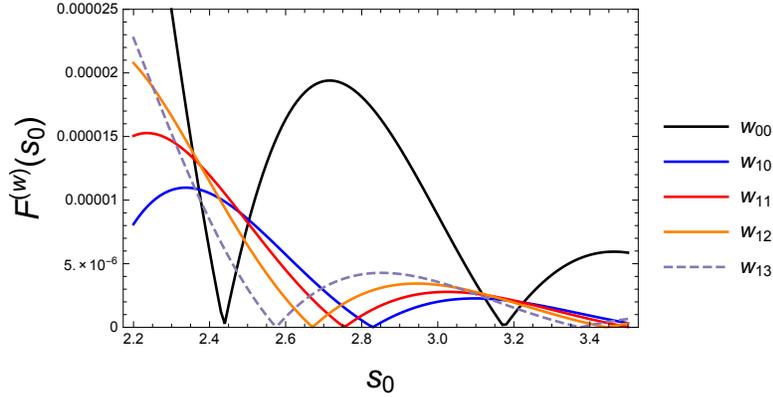,width=4in}}
\vspace*{8pt}
\caption{Contribution from DVs to the FESR for the 5 pinched weights of the old-strategy method, Eq. (\ref{demystifying}).\protect\label{fig7}}
\end{figure}

Fig.~\ref{fig7} shows the result of the integral
\begin{equation}
\label{demystifying}
F^{(w_{kl})}(s_0)=\left|\int_{s_0}^{\infty}\frac{ds}{s_0} \ w_{kl}(\frac{s}{s_0}) \ \frac{1}{\pi}\mathrm{Im}\Pi^V_{DV}(s)\right|
\end{equation}
as a function of $s_0$ and the five pinched weights employed in the old-strategy analysis\cite{oldstrategy,Davier} for the values of the DV parameters found in the analysis of Ref.~\refcite{us}. Specifically, the parameters used are $\delta_V\simeq 3.5, \gamma_V\simeq 0.62\, \mathrm{GeV}^{-2}, \alpha_V\simeq -2.43$ and $\beta_V\simeq 4.32\,\mathrm{GeV}^{-2}$, corresponding to the $s_{\mathrm{min}}=1.55\ \mathrm{GeV}^2$, FOPT result of Table I in Ref.~\refcite{us}.\footnote{See Ref.~\refcite{Boito0} for the precise definition of FOPT and CIPT.} Although $w_{00}$ is only doubly pinched, whereas all the other $w_{kl}$ are triply pinched (see section 5), Fig.~\ref{fig7} shows explicitly how all the triply pinched moments actually produce a \emph{larger} contribution to the FESR from DVs than the doubly pinched weight $w_{00}$ near the value of $s_0=m_{\tau}^2\simeq 3.16\ \mathrm{GeV}^2$ where the old-strategy analysis is made. This is of course against the naive belief that more pinching reduces DVs more, and  is a consequence of the DV function in Eq.~ (\ref{DV}) not having a definite sign. On the other hand, when a window in $s_0$ is considered, one sees in Fig. ~\ref{fig7} how the amplitude of the oscillations for $w_{00}$ is larger than those for the triply pinched weights, as one would naively have guessed. So, again, the lesson is that considering a window in $s_0$ is safer than a just a single value, even if this is the largest possible, $s_0=m_{\tau}^2$.

\section{An illustrative exercise: Hunting for systematic errors.}

Even though neglecting DVs by fiat is not the best way to quantify their effect on $\alpha_s$, one might be tempted to think that, if the weight is sufficiently pinched and the FESR shows a good match between the fit and the data as $s_0$ is lowered from $m_{\tau}^2$, the analysis is safe in the sense that systematic effects caused by non-vanishing DVs are numerically negligible, in particular for the $V+A$ channel since the oscillations in the spectrum are smaller there.

\begin{figure}[t]
\centerline{\psfig{file=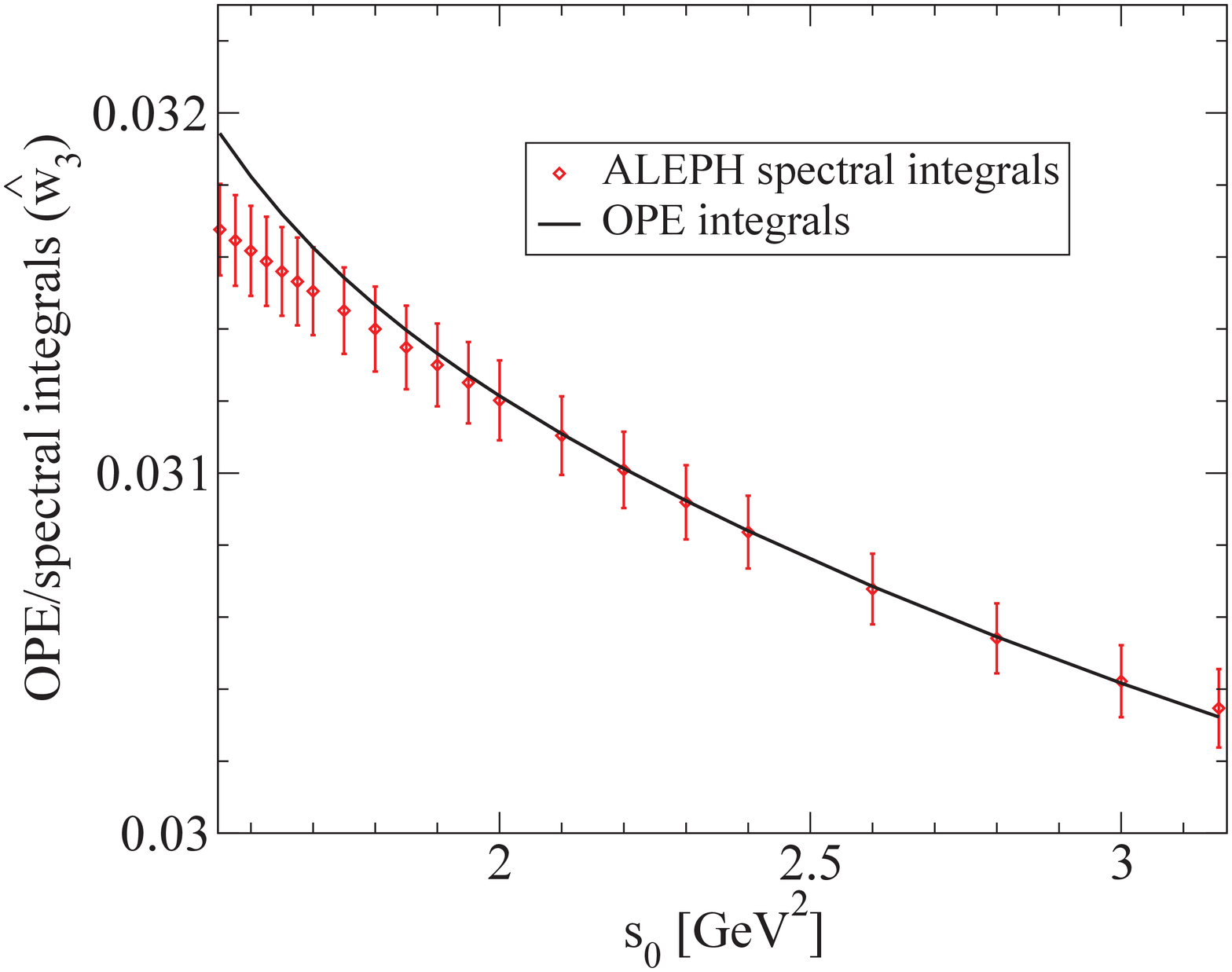,width=2.5in}}
\vspace*{8pt}
\caption{Comparison of the $w_{00}$-weighted spectral integral and the OPE integral evaluated with the no-DV fit given in Eq.~(\ref{noDVfit}).\protect\label{fig8}}
\end{figure}

Fig. ~\ref{fig8} shows the result of a fit to $V+A$, FOPT,  for $w_{00}$ (i.e., the same weight as in the decay width) for $1.95\ \mathrm{GeV}^2\leq s\leq m_{\tau}^2$, \emph{without} DVs. It is necessary to fit in a window of $s_0$ because the FESR depends on three parameters: $\alpha_s, C_{6,V+A}$  and $C_{8,V+A}$. The optimal fit, with a high p value of $57\%$,  is obtained using the fit window $2.2\, \mathrm{GeV}^2 < s_0 < m_{\tau}^2$ and yields
\begin{eqnarray}
\label{noDVfit}
\alpha_s(m_\tau^2)&=&0.330\pm 0.006\ ,\nonumber\\
C_{6,V+A}&=&0.0070\pm 0.0022\ {\rm GeV}^6\ ,\nonumber\\
C_{8,V+A}&=&-0.0088\pm 0.0042\ {\rm GeV}^8\ .
\end{eqnarray}
As one can see, the quality of the fit is excellent and it would make nobody suspect that there is a systematic effect lurking in the results. However, it is there.

\begin{figure}[t]
\centerline{\psfig{file=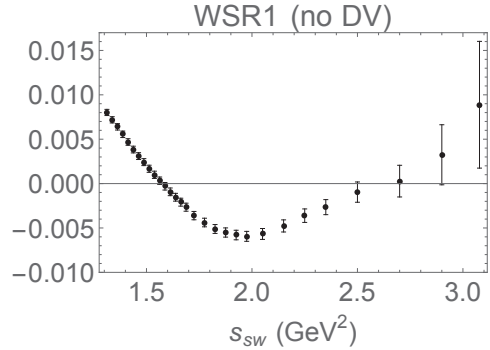,width=2.5in}}
\vspace*{8pt}
\caption{The 1st Weinberg sum rule for the no-DV fit of Eq.~(\ref{noDVfit}).\protect\label{fig9}}
\end{figure}

To unravel this systematic effect, one may look at the first Weinberg sum rule (WSR) for this case. This is shown in Fig.~\ref{fig9}. The value of $s_{sw}$ shown in abscissa is the point up to which the integral in the WSR  is taken. Clearly the WSR is violated even at those $s_0$ values for which there is a good match in Fig.~\ref{fig8}. Since the V-A spectral function entering the WSR vanishes to a high level of approximation in the OPE, the non-zero result in Fig.~\ref{fig9} is only due to DVs.

\begin{figure}[t]
\centerline{\psfig{file=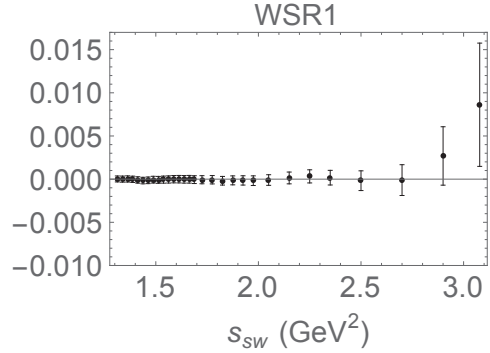,width=2.5in}}
\vspace*{8pt}
\caption{The 1st Weiberg sum rule from Eq.~(\ref{DVfit}) and the results from Ref.~\protect\refcite{us}.\label{fig10}}
\end{figure}

To be able to assess in more quantitative terms the systematic effect on $\alpha_s$ which, after all, is the main objective we are after, one may repeat the fit leading to Fig.~\ref{fig8}, but now taking as external input the result of the DV parameters obtained in Ref.~\refcite{us} from the $s_{\mathrm{min}}=1.55\ \mathrm{GeV}^2$ fit of Table V, and fitting the OPE parameters,  $C_{(6,8),V+A}$ and $\alpha_s$. The change is dramatic. The new values obtained are now

\begin{eqnarray}
\label{DVfit}
\alpha_s(m_\tau^2)&=&0.301\pm 0.006\pm 0.009\ ,\nonumber\\
C_{6,V+A}&=&-0.0127\pm 0.0020\pm 0.0066\ {\rm GeV}^6\ ,\nonumber\\
C_{8,V+A}&=&0.0399\pm 0.0040\pm0.021\ {\rm GeV}^8\ ,
\end{eqnarray}
instead of Eq.~(\ref{noDVfit}), where the first error is statistical and the second is the one induced by the correlations with the external input DV parameters. As one can see, the condensates even flip sign relative to Eq.~(\ref{noDVfit}), and the value of $\alpha_s(m_\tau^2)$ goes \emph{down} by a lot more than the nominal error obtained in the fit (\ref{noDVfit}). Furthermore, unlike the previous case of the WSR in which DVs were neglected (Fig.~\ref{fig9}), now the WSR looks like Fig.~\ref{fig10}, where $s_{sw}$ is the point at which one switches from the spectral data to the DV parametrization in the infinite integration interval of the WSR. This concludes the exercise.

\section{Results}

Let us now compare the final results obtained for $\alpha_s$ from $\tau$ decays with and without including the DV parametrization  (\ref{DV}). With DVs included, we obtain\cite{us}
\begin{eqnarray}
\label{results}
 \alpha_s(m_{\tau}^2)&=&{0.296\pm 0.010} \  \longrightarrow \ \alpha_s(m_{Z}^2)={ 0.1155\pm 0.0014}\quad (\mathrm{FOPT}) \nonumber \\
 \alpha_s(m_{\tau}^2)&=&{ 0.310\pm 0.014}\   \longrightarrow \ \alpha_s(m_{Z}^2)={ 0.1174\pm 0.0019}\quad (\mathrm{CIPT})\ ,
 \end{eqnarray}
while the old-strategy method produces a shift of the order $+0.03$ higher  in $\alpha_s$ at the $\tau$ mass and with errors which are about half those shown in Eq.~(\ref{results}).\cite{Davier} The systematic increase in the value of $\alpha_s$ is of course reminiscent of the results found in the exercise of the previous section.

It is sometimes customary to express the result in terms of the $R_{V+A}$ ratio in Eq. ~(\ref{R}):
\begin{equation}
R_{V+A}=N_c\ S_{EW} |V_{ud}|^2\Big( 1+ \delta_P + \delta_6 + \delta_8 + \delta_{DV}\Big)\ ,
\end{equation}
splitting the QCD contributions into a perturbative term $\delta_P$ and a nonperturbative term $\delta_{NP}=\delta_6+\delta_8+\delta_{DV}$, where $\delta_{6,8}$ are the contributions from the corresponding dimension in the OPE and $\delta_{DV}$ are the DV contributions. In this case we obtain\cite{us}
\begin{eqnarray}
\label{deltas}
\delta_{NP}&=&0.020 \pm 0.009\quad (\mathrm{FOPT})\nonumber \\
\delta_{NP}&=&0.016 \pm 0.010\quad (\mathrm{CIPT})\ ,
\end{eqnarray}
to be compared to the value obtained in the old-strategy method, $\delta_{NP}=-0.0064 \pm 0.0013\   (\mathrm{CIPT})$,\cite{Davier} again with errors which are claimed to be much smaller than those in Eq.~(\ref{deltas}). Since the results of the old-strategy method did not consider DVs and neglected higher-dimension condensates, these errors are to be considered an underestimate.

Finally, since perturbative and non-perturbative contributions are strongly correlated (since their sum has to equal the spectral function integral) we would like to warn about studies of the perturbative term $\delta_P$ which use a previous result for $\delta_{NP}$ as if they were independent of each other. For any new proposal different from CIPT or FOPT to reorganize pertubation theory, a new $\delta_{NP}$ should  be obtained from a new fit to the FESRs.

\section{Conclusions and Outlook}

The spectral functions in Fig.~\ref{fig1} show visible oscillations which are due to DVs i.e., terms which go beyond the OPE. DVs, therefore, are not just a question of principle but they also exist in practice. How much DVs impact a determination of $\alpha_s(m_{\tau}^2)$ cannot be answered by neglecting DVs a priori. Pinching may help suppress DVs (especially if an interval in $s_0$ is used in the analysis) but one should remember that it is not possible to simultaneously reduce contributions from DVs and the higher-dimension condensates. So, there is a price to pay. Ignoring DVs and/or the contribution from these higher-dimension condensates introduces an unquantified systematic error that has plagued the old-strategy method.\cite{oldstrategy,Davier}

On the other hand, the use of a concrete physically motivated parametrization of DVs allows one to perform a complete reanalysis from scratch and quantitatively check the reliability of all the assumptions made  in the old-strategy method. Doing so has unravelled significant systematic effects, the most important of which is a positive shift in $\alpha_s(m_{\tau}^2)$ of order  $+0.03$ when the old strategy is employed. Therefore, any error claimed smaller than $0.03$ in $\alpha_s(m_{\tau}^2)$ from any previous analysis of hadronic $\tau$ decays neglecting DVs must be considered an underestimate. In order to quantify potential systematic errors a concrete parametrization of DVs for QCD must be assumed.

For our parametrization of DVs in Eq.~(\ref{DV}), since DVs cannot yet be derived from QCD from first principles, we have used arguments as general  as possible to minimize model dependence. In this sense it is important to realize that the old-strategy method is not free of model dependence.  This model dependence can be summarized by the following arbitrary choices made there: $\mathrm{e}^{-\delta_{V/A}}=0$ (in Eq.~(\ref{DV})) and $C_{10,12,14,16}=0$, where the $C_n$'s are the corresponding OPE condensates. None of these choices are favored by present data or theoretical knowledge. Unlike the case of the old-strategy method, in our case DVs are determined by the data and the analysis passes all known tests up to date, with better performance than the old-strategy analysis.

At this point it is difficult to make progress without a better theoretical understanding of DVs. In this sense, ideas like resurgence\cite{Shifman2} or functional analysis methods\cite{Caprini} may be helpful. It would also help to have more detailed inclusive experimental spectral data, such as the one which could, in principle, be obtained from the BABAR and Belle experiments. We would like to encourage our experimental colleagues to provide us with this important information in the near future.

\vspace{1.5cm}

\textbf{\underline{Note added}}

After completion of this writeup, a new analysis using the ``old
strategy'' appeared.\cite{BS}  Here we just note that this new analysis
fails to answer most of the criticisms of the old strategy raised in Sec. VII of Ref.~\refcite{us}, and which are reviewed here.  We also point out that even though Ref.~\refcite{BS} dismisses the use of ansatz (\ref{DV}) in the
determination of $\alpha_s$ from $V+A$, the authors advocate the use of precisely this type of ansatz in the extraction of
low-energy constants and OPE condensates from $V-A$  in Refs.~\refcite{Gonzalez} and \refcite{Rodriguez}.

\vspace{.5cm}

\section*{Acknowledgments}

We would like to thank the Mainz Institute for Theoretical Physics and, in particular, the organizers of this workshop for creating a very fruitful atmosphere.

MG is supported in part by the US Department of Energy under contract
DE-FG02-92ER40711, and DB's work is supported by the S\~ao Paulo Research Foundation
(FAPESP), grant 15/20689-9, and by CNPq, grant 305431/2015-3.
SP is supported by CICYTFEDER-FPA2014-55613-P, 2014-SGR-1450.
KM is supported by a grant from the Natural Sciences and
Engineering Research Council of Canada.

\end{document}